\documentclass[reprint, 10pt, floatfix,aps,twocolumn,a4paper,superscriptaddress,nofootinbib,notitlepage,pra,letterpaper]{revtex4-1}
\usepackage{inputenc}
\usepackage[T1]{fontenc} 
\usepackage{paralist}
\usepackage{hyperref}
\usepackage{amsopn,amsthm,dsfont,amssymb}
\usepackage{mathrsfs}
\usepackage{cancel} 
\usepackage{color} 
\usepackage{soul} 
\usepackage[normalem]{ulem}
\usepackage{cancel} 
\usepackage{hyperref} 

\usepackage{soul,xcolor}
\usepackage{lipsum}
\usepackage[english]{babel}
\usepackage{comment}
\hypersetup{
    colorlinks=true,
    linkcolor=blue,
    filecolor=magenta,      
    urlcolor=cyan,
}

\AtBeginDocument{%
    \newwrite\bibnotes
    \def\bibnotesext{Notes.bib}
    \immediate\openout\bibnotes=\jobname\bibnotesext
    \immediate\write\bibnotes{@CONTROL{REVTEX41Control}}
    \immediate\write\bibnotes{@CONTROL{%
    apsrev41Control,author="08",editor="1",pages="1",title="0",year="1"}}
     \if@filesw
     \immediate\write\@auxout{\string\citation{apsrev41Control}}%
    \fi
}%

\newcommand{\ket}[1]{|#1\rangle}
\newcommand{\bra}[1]{\langle #1|}

\usepackage{mathtools}

\usepackage{adjustbox}
\usepackage[caption=false]{subfig}

\begin{document}
\title{Testing Gravitationally-Induced-Entanglement with Massive Particles and Quantum Clocks }
\author{Germain Tobar}
\affiliation{Department of Physics, Stockholm University, SE-106 91 Stockholm, Sweden}

\author{Markus Aspelmeyer}
\affiliation{University of Vienna, Faculty of Physics, Vienna Center for Quantum Science and Technology (VCQ),
Vienna, Austria}
\affiliation{Institute for Quantum Optics and Quantum Information (IQOQI) Vienna, Austrian Academy of Sciences,
Vienna, Austria}

\author{Magdalena Zych}
\affiliation{Department of Physics, Stockholm University, SE-106 91 Stockholm, Sweden}

\begin{abstract}

We present an experimental proposal to test whether gravity can entangle a levitated particle with an array of quantum clocks. By harnessing the collective interaction between a micrometer-scale particle and an array of ultra-precise nuclear clocks, our scheme offers a conceivable path to observing gravitationally mediated entanglement with quantum clocks in the laboratory. Crucially, this setup will probe a genuinely relativistic effect one that requires the contribution of internal clock energy to active gravitational mass and cannot arise from the Newtonian interaction between fixed masses. Therefore, observing this entanglement would provide evidence for the quantised nature of active gravitational mass -- which has not been tested in any prior experiment -- rather than for spatial coherence of fixed masses,which is routinely tested in matterwave interferometers.\\

\end{abstract}

\maketitle
\textit{Introduction.}
Determining whether gravity is fundamentally quantum remains one of the most important open questions in physics \cite{Carney_2019,RevModPhys.97.015003}. This challenge has sparked on-going debate: quantum gravity theories and quantum information based programs pursue the quantisation of general relativity \cite{alma991014454786403131} or distinctive features such as its causal structure \cite{OreshkovOgnyan2012Qcwn,PhysRevA.88.022318}. Other approaches propose that gravity should remain fully classical —requiring modifications to quantum mechanics at gravitational scales \cite{CarlessoMatteo2022Psaf} or modifying gravity in such a way to mimic some quantum-like effects through stochastic or non-local models. Resolving this question requires experimental tests whose results cannot be explained with classical gravity. Current experimental efforts typically follow one of two approaches: probing quantum features of massive sources of gravity \cite{bose2017spin,MVedral,zehfeynman,hl1c-t8z9}, or testing quantum properties of gravitational radiation \cite{blencowe2013effective,PhysRevD.98.124006,parikh2020noise,kanno2021noise,Guerreiro2022quantumsignaturesin,tobar2023detecting,manikandan2024detectingacoherenceradiationfields}.

The idea of testing quantum features of the gravitational field of non-classical states of source masses dates back to Feynman’s influential remarks at the Chapel Hill conference \cite{CécileM.DeWitted.2011TRoG}, in which it was first discussed whether gravitational objects can exhibit quantum coherence and entanglement. More recently, this idea has been revisited with proposals to test entanglement between two massive objects \cite{bose2017spin,MVedral,MarshmanGIE}. Consequently, the correlations alone do not isolate a relativistic gravitational degree of freedom and may admit non-local classical accounts \cite{PhysRevD.108.L101702}. Moreover, these experiments can often be described entirely using Newtonian gravity \cite{fragkos2025probingclassicalquantumviolations, Foo2026}. Related work has identified signatures of delocalised sources that go beyond the Newtonian potential \cite{hl1c-t8z9}. Fianlly, a pre-dating proposal \cite{RuizEstebanCastro2017Eoqc} examined the quantum contribution of internal mass-energy to active gravitational mass, predicting gravity-generated entanglement between quantum clocks; however, its predicted timescales remain far beyond present capabilities.

In this Letter, we propose a new experimental platform that combines the advantages of both approaches: a levitated massive particle interacting gravitationally with an array of quantum clocks. This setup is motivated by finding a scalable test of gravitationally induced entanglement and decoherence in a regime where relativistic features of gravity are essential. By leveraging both the large mass of the particle and the collective enhancement from multiple clocks, we demonstrate that relativistic quantum effects of gravity can become observable within the capabilities of future experiments, overturning current pessimistic estimates.

\begin{figure*}[t]
\begin{center}
\includegraphics[scale  = 1.2]{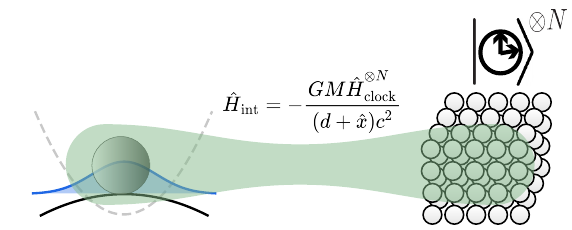}
    \caption{\label{levitatedlock} The gravitational interaction between an array of quantum clocks and a macroscopic levitated particle in a delocalised quantum state, enhances the gravitationally induced entanglement rate by both the macroscopic number of quantum clocks $N$ and the mass of the levitated particle $M$. This combined enhancement is sufficient to enable a test of gravitationally-induced-entanglement with quantum clocks that is within reach of realistic experimental capabilities.}
\end{center}
\end{figure*}

\textit{Gravitational interaction between two clocks.} It is straightforward to see the challenge of testing gravitationally mediated entanglement between quantum clocks by considering the extremely weak nature of the gravitational interaction between them. Consider gravitational interaction between two particles of masses $m_A$ and $m_B$ separated by a distance $x$, $\hat{H}_\mathrm{int} = \frac{G}{c^4 x} m_A m_B$. If we now invoke the mass-energy equivalence and replace \cite{ZychMagdalena2018Qfot,PhysRevA.106.052801,PhysRevLett.134.093201}, $m_{A} \rightarrow m_A + \hat{H}_A/c^2$ and $m_{B} \rightarrow m_B + \hat{H}_B/c^2$, the gravitational interaction between two clocks reads
\begin{equation}\label{hnit}
\hat{H}_\mathrm{int} = \frac{G}{c^4 x} \hat{H}_A \hat{H}_B,
\end{equation}
where $\hat{H}_{A(B)}$
is the rest frame internal Hamiltonian of clock $A (B)$, respectively. When both clocks are initialized in the superposition of their ground $\ket{g}$ and excited $\ket{e}$ state, $\ket{+} = \frac{1}{\sqrt{2}}(\ket{g} + \ket{e})$ so that the joint initial state is $\ket{\psi(0)} = \ket{+}\ket{+}$, it was shown in Ref.~\cite{RuizEstebanCastro2017Eoqc} that the Hamiltonian in Eq.~\eqref{hnit} generates entanglement (and thus decoherence of the individual subsystems of an initially pure bipartite state, absent other decoherence sources serves as a signature of entanglement) on a time-scale
\begin{equation}
    t_{\mathrm{clocks}} = \frac{2c^4 x}{\hbar G \omega^2}.    
\end{equation}
It is the presence of a factor of $(\omega / c^2)^2$, which makes these effects practically unobservable: for typical optical clock frequencies $\omega_c / 2\pi \sim 10^{14} \; \mathrm{Hz}$, and assuming their separation is $x = 1 \; \mathrm{\mu}m$, $t_{\mathrm{clocks}} \sim 10^{41} \; \mathrm{s}$—an astronomically long duration. Replacing the two clocks with an array of a macroscopic number ($N \sim 10^{23}$) of optical clocks as suggested in Ref.~\cite{RuizEstebanCastro2017Eoqc} lowers this time scale to $t_{\mathrm{clocks}} \sim 10^{31}$s, still beyond the lifetime of our visible universe.

The possibility to test gravity-mediated entanglement can be enhanced by replacing one of the clocks with the macroscopic mass of a levitated particle. Let us first look at the interaction between a particle with mass  $M$  and a single clock. The the interaction Hamiltonian reads
\begin{equation}\label{clockH}
    \hat{H}_\mathrm{int}\left(\hat{x}_c, \hat{x}_p\right)=-\frac{G M}{\left|\hat{x}_c-\hat{x}_p\right|} \frac{\hat{H}_{\text {clock }}}{c^2},
\end{equation}
where $\hat{x}_c$ and $\hat{x}_p$ are the positions of the clock and the levitated particle respectively as in the previous paragraph $\hat{H}_{\text {clock }} = \frac{\hbar\omega_c\hat{\sigma}_z }{2}$ is the clock Hamiltonian.

Assuming the levitated particle is in a spatial superposition state $\ket{\psi_p} = \frac{1}{\sqrt{2}}\left(\ket{-x} + \ket{+x} \right)$, with delocalisation $\Delta x$ (Fig.~\ref{levitatedlock}), the clock state becomes entangled with the position of the particle and $|\psi(t)\rangle=\frac{1}{\sqrt{2}}\left(|-x\rangle\left|\tau_{-}\right\rangle+|+x\rangle\left|\tau_{+}\right\rangle\right)$, producing decoherence of either the clock state or the spatial coherence of the particle on the time-scale (see Appendix~\ref{decapp1})
\begin{equation}\label{tdm}
   t_{\mathrm{mass-clock}}^{{\mathrm{\Delta x\gg d}}}  =  \frac{2c^2d}{GM\omega_c}.
\end{equation}
Here, d is the distance from the clock to the closer branch of the massive particle's superposition. In the regime where $\Delta x \gg d$, assuming a Planck mass levitated particle $M\sim 10^{-8} \; \mathrm{kg}$, $d = 100 \; \mathrm{\mu m}$, $\omega_c / 2 \pi \sim 10^{14} \mathrm{~Hz}$, this decoherence time-scale shortens to $t_{\mathrm{mass-clock}} \sim 10^{16} \; \mathrm{s}$, an improvement by more than 15 orders of magnitude compared to the two-clocks scenarios above. Alternatively, in the regime for which the massive particle's superposition  is much smaller than the interaction distance $\Delta x \ll d$ (see Appendix~\ref{smalldelocalisation}), the decoherence time is longer by $d/\Delta x$,  it reads
\begin{equation}\label{tdm-smallDeltax}
   t_{\mathrm{mass-clock}}^{{\mathrm{\Delta x\ll d}}} = \frac{2c^2d^2}{GM \omega_c \Delta x}.
\end{equation}
\textit{Massive particle interacting with N-clocks.} We now move to the central proposal of this work: how the combination of the previously described enhancement due to the macroscopic mass of the levitated particle with the enhancement due to a macroscopic number of quantum clocks explored in Ref.~\cite{RuizEstebanCastro2017Eoqc} enables a conceivable test of gravitationally mediated entanglement beyond the non-relativistic limit. For the massive particle interacting with $N$ clocks, the decoherence time-scales in Eq.~\eqref{tdm} and \eqref{tdm-smallDeltax} are reduced by a factor of $\sqrt{N}$ (see Appendix~\ref{decapp1}). In the large superposition regime we thus have
\begin{equation}\label{td1}
    t_{\mathrm{mass}-{N}-{clocks}}^{{\mathrm{\Delta x\gg d}}}=\frac{2c^2d }{G M\sqrt{N}\omega_c}. 
\end{equation}
For Avogadro number of clocks , $N = 10^{23}$, the previous estimates are thus reduced by a factor $\simeq10^{12}$ and decoherence time scale becomes $t_{\mathrm{mass}-{N}-{clocks}} \sim 10^{4} \mathrm{~s}\sim 17 \mathrm{~h}$. 
Most importantly, for $N = 10^{23}$ nuclear clocks with transition frequencies on the order of $\omega_c/2\pi \sim 10^{18} \; \mathrm{Hz}$, available in scandium \cite{Shvyd}, we obtain a decoherence timescale $t_{\mathrm{mass}-{N}-{clocks}} \sim 1 \; \mathrm{s}$. 
Such a short time warrants considering the experimentally relevant regime of Gaussian particle wave packets with small delocalisation $\Delta x$. For $\Delta x \ll d$ we have (See Appendix~\ref{smalldelocalisation})
\begin{equation}\label{tau1234}
  t_{\mathrm{mass}-{N}-{clocks}}^{{\mathrm{\Delta x\ll d}}} = \frac{2 c^2 d^2}{G M\sqrt{N} \omega_c\Delta x},
\end{equation}
which for a Planck-mass particle and Avogadro number of clocks with scandium clock transition ($d=450\,\mu\mathrm{m}$, $\Delta x=0.4\,\mu\mathrm{m}$) yields $t_{\mathrm{mass}-{N}-{clocks}}\sim 5\cdot 10^{3}\,\mathrm{s}\sim 2\,\mathrm{h}$. In the more realistic situation in which the levitated particle interacts with a crystal slab containing clocks of density $n$, we obtain (see Appendix~\ref{crystal123}):
\begin{equation}\label{eq:tfinite}
\begin{split}
        t_{\text {finite }} &\approx \frac{2 c^2}{G M_p \omega_c \Delta x} \sqrt{\frac{2 D}{\pi n}},\\
        &\approx \frac{2 c^2}{G M_p \omega_c \Delta x} \sqrt{\frac{2 DV}{\pi N}},
\end{split}
\end{equation}
where $D$ is the distance from the levitated particle to the closest face of the crystal slab, $V$ is the volume of the crystal and $N$ the number of clocks. 
Taking the parameters $D=450\,\mathrm{\mu m}$, $M_p = 2.18\times10^{-8}\,\mathrm{kg}$, $\omega_c = 1.82\times10^{19}\,\mathrm{s^{-1}}$, $\Delta x = 0.1\,\mathrm{\mu m}$ (corresponding to an aspirational but realistic regime of de-localisation in Ref.~\cite{PhysRevLett.127.023601}), $n = \frac{N}{V} = 4\times10^{28}\,\mathrm{m^{-3}}$ for a centimetre-scale crystal, the conditions $\Delta x < D < V^{\frac{1}{3}}$ are satisfied. Substituting these values gives (for $10^{23}$ clocks),
\begin{equation}\label{eq:tfinite-estimate}
    t_{\mathrm{finite}}\approx 5\times10^6\,\mathrm{s}
    \approx 64\,\mathrm{Days}.
\end{equation}
Equation~\eqref{eq:tfinite} introduces a correction to Eq.~\eqref{tau1234} that accounts for the distribution of clocks within a finite volume, replacing $d^2$ with $\sqrt{\frac{2DV}{\pi}}$. We use this corrected expression in the subsequent analysis. We remind the reader that this decoherence scale concerns both, the massive particle and any of the individual clocks. Focusing on the clocks, resolving such a decoherence rate above a typical $\mathrm{Hz}$-scale linewidth in scandium would require $10^{10}$ measurements, which can be accessible through parallel independent measurements across a fraction of the $10^{23}$ clocks in the crystal.

\textit{Dynamics.} We now consider the complete dynamics of N clocks gravitationally interacting with a levitated particle, justifying the use of Eq.~\eqref{clockH} as the core model for our predictions. We begin with the Hamiltonian of $N$ mutually non-interacting quantum clocks modelled as particles trapped in a potential, gravitationally interacting with a levitated particle. Due to the high frequency of the internal clock states, we consider that the quantised mass-energy \cite{ZychMagdalena2018Qfot} only contributes to the clocks, while neglecting the quantised mass-energy corrections of the levitated particle. In this case the Hamiltonian of the array of clocks trapped against Earth gravity by a potential $V(\cdot)$ and interacting with the particle is
\begin{equation}\label{Htot2}
\begin{split}
    \hat{H} &= \frac{\hat{p}_p^2}{2M_p} + \sum^N_{i = 1}\left( \hat{H}_{\text {clock }}^i + \frac{(\hat{p}^{i}_\mathrm{clock})^2 }{2 \hat{M}_\mathrm{clock}^i} + V(\hat{x}_\mathrm{clock}^i)\right)+ \\
    &  -\frac{G M_p}{\left|\hat{x}_\mathrm{clock}^i-\hat{x}_p\right|} \frac{\hat{H}_{\text {clock }}}{c^2}.
\end{split}
\end{equation}
We can directly apply the results of Ref.~\cite{haustein2019massenergyequivalenceharmonicallytrapped} in order to compare the orders of magnitude of the mass-energy equivalence induced phase shifts, frequency shifts and visibility modulation of the clocks. We first discuss the effects arising due to Earth's gravitational background before focusing on the effects due to the levitated particle. 

The model captures the fact that each clock experiences a fractional frequency shift due to the combination of the second order Doppler shift and a gravitational shift due to mass-energy dependent displacement of the total trap minimum. For scandium atoms confined in a crystal lattice, the effective trap frequency is $\omega_t\approx 10^{13}$Hz, \cite{PhysRev.175.782} and the two shifts are estimated as $\delta_{2D}\approx10^{-13}$ and $\delta_g\approx10^{-43}$, respectively. Note that for current state-of-the-art clocks the fractional frequency precision is on the level of $10^{-18}-10^{-19}$, \cite{arnold2026lu, Fortier:26}. The Doppler shift in these clocks is already of the same order of magnitude, but it is a systematic which can be eliminated. We thus consider that similarly for scandium nuclear clocks the $\delta_{2D}$ can be accounted for among systematics while the shift $\delta_g$ can be neglected. The position and motional couplings result also in a coherence modulation of the clock states. For the initial sate of the scandium atom approximated by its motional ground state, the Ramsey visibility is modulated only at the level of $10^{-14}$.

We now discuss analytical description of the dynamics under Eq.~\eqref{Htot2}, for which we expand the quantised position of the levitated particle about c-number positions $x_L^M$ and $x_R^M$ corresponding to the eigenvalues of the semi-classically localised states $\ket{x_L^M}$ and $\ket{x_R^M}$ in the left and right branches of the superposition, respectively. The Hamiltonian now becomes:
\begin{equation}\label{Hhere}
\begin{split}
    \hat{H} &= \sum_i  \hat{H}_{\text {clock }}^i+\frac{\hat{p}_i^2}{2\hat{M}_{\text {clock }}^i} + V\left(\hat{x}_{\text {clock }}^i\right) \\ &- \frac{GM_p}{d + x_L}\left(1 - \frac{\hat{x}_m^i}{d + x_L} \right)\frac{\hat{H}_{\text {clock }}}{c^2}\ket{x_L^M}\bra{x_L^M} \\
    &- \frac{GM_p}{d + x_R}\left(1 - \frac{\hat{x}_m^i}{d + x_R} \right)\frac{\hat{H}_{\text {clock }}}{c^2}\ket{x_R^M}\bra{x_R^M},
\end{split}
\end{equation}
where  $\hat{x}_m^i$, corresponds to the position of the $i$-th clock from the centre of its confining potential. Assuming again a harmonic trap for each clock, we block-diagonalise the Hamiltonian in a similiar manner to that in Ref.~\cite{haustein2019massenergyequivalenceharmonicallytrapped}. 
For each internal energy state, of the clock, there is a harmonic oscillator with mass $M_k = m(1 + \frac{E_k}{m c^2})$, and spatial co-ordinate that diagonalises Eq.~\eqref{Hhere},
\begin{equation}
\hat{x}_{g,\alpha,k}:=\hat{x}+g_\alpha / \omega_k^2,  
\end{equation}
with re-defined gravitational accelaration
\begin{equation}
    g_\alpha=g_E+\frac{G M_p}{\left(d+x_\alpha\right)^2},
\end{equation}
where $g_E$ is the Earth's gravitational acceleration.

Introducing orthonormal projectors onto the internal energy states of each clock $\hat{\Pi}_k:=\left|E_k\right\rangle\left\langle E_k\right|$, the Hamiltonian of Eq.~\eqref{Hhere} can thus be diagonalised for each individual clock as 
\begin{equation}\label{Hpix}
     \hat{H}^i =  \sum_{k,\alpha} \hat{h}_{i,k,\alpha} \otimes \hat{\Pi}_k \otimes \left|x_\alpha^M\right\rangle\left\langle x_\alpha^M\right|,
\end{equation}
where 
\begin{equation}
\begin{split}
     \hat{h}_{i, k ,\alpha} &= M_{i, k} c^2\left(1-\frac{GM_p}{d + x_\alpha}-\frac{g_\alpha^2}{2 \omega_k^2 c^2}\right) + \hbar \omega_k\left(\hat{n}_k+\frac{1}{2}\right), 
\end{split}
\end{equation}
and where, $\omega_k:=\left(\kappa / M_k\right)^{1 / 2}$, such that $\kappa$ is the harmonic trap stiffness. The number operators are defined through the annihilation operators 
\begin{equation}
    \hat{a}_{g,\alpha,k} =\sqrt{\frac{M_k \omega_k}{2 \hbar}}\left(\hat{x}_{g,\alpha,k}+\frac{i \hat{p}}{M_k \omega_k}\right).
\end{equation}

The eigenbasis of the Hamiltonian is of the form $\left\{\left|E_k\right\rangle,\left|n_{k,\alpha}\right\rangle,\left|x_\alpha^M\right\rangle\right\}_{k\in\mathbb{N},\,\alpha\in\{L,R\}}$. For the scandium clock parameters and deep crystal trap discussed above, the single branch energy difference, up to order $1/c^2$, reads 
\begin{equation}
\begin{aligned}
E_{k,\alpha,n}-E_{0,\alpha,n}
\simeq&\\
\Delta E_{k0}
&\left(\!1-\frac{\hbar\omega_0}{2M_0c^2}\!\left(n+\frac12\right)\!
-\frac{GM_P}{c^2R_\alpha}
-\frac{g_\alpha^2}{\omega_0^2c^2}\!\right).
\end{aligned}
\end{equation}
The first term is the gravitational redshift due to the levitated particle, the second term (proportional to the trap frequency $\hbar\omega_0$) is the second-order-Doppler shift, and the last term is the correction due to the change of the minimum of the trap due to gravity from the levitated particle.   For our choice of parameters $M_P=2.18\times10^{-8}\,\mathrm{kg}$, $R_L=D=450\,\mu\mathrm{m}$, and $R_R-R_D=\Delta x=0.1\,\mu\mathrm{m}$, the second-order-Doppler shift
is the same as already estimated (and it is common-mode for each branch of the levitated particle superposition).  The correction due to gravity-dependence of the trap minimum is negligibly small, with the associated  contrast modulation  on the order of $10^{-70}$ in each branch. The dominant effect resulting from the particle on the clocks is its gravitational redshift, and we next derive the arising entanglement.  

\begin{table}[t]
\begin{ruledtabular}
\begin{tabular}{lcccc}
Clock Type & Number of clocks & Decoherence rate \\
\hline
Atomic & $10^6$ & $10^{-10} \; \mathrm{Hz}$   \\
Nuclear Sc-45 & $10^6$ & $10^{-6} \; \mathrm{Hz}$  \\
Atomic & $10^{23}$ & $10^{-10} \; \mathrm{Hz}$  \\
Nuclear Sc-45 & $10^{23}$ &  $10^{-6} \; \mathrm{Hz}$ 
\end{tabular}
\end{ruledtabular}
\end{table}

With the Hamiltonian in diagonal form of Eq.~\eqref{Hpix}, we consider the dynamics of an array of clocks, with the levitated particle in the previously described cat state, and we approximate the centre of mass of each clock in the ground state $\ket{g}^{\otimes N}$ by
\begin{equation}
    \ket{\psi(0)} = \frac{1}{\sqrt{2}}\left(\ket{x_L} +\ket{x_R}\right)\ket{+}^{\otimes N}\ket{g}^{\otimes N}.
\end{equation}
After letting the systems interact for time $\tau$, then recombining the wave-packet of the particle and projecting onto $\ket{\pm_P} =  \frac{1}{\sqrt{2}}\left(\ket{x_L} \pm \ket{x_R}\right)$, the clocks are left in the many-body entangled state
\begin{equation}
\begin{split}
    \ket{\psi(\tau)} &= \sum_k \frac{1}{\sqrt{2}}(\exp(-i\tau \hat{h}_{k, L} \otimes \hat{\Pi}_k)^N\ket{+}^{\otimes N}\ket{g}^{\otimes N} \\&+  \exp(-i\tau \hat{h}_{k, R} \otimes \hat{\Pi}_k)^N\ket{+}^{\otimes N}\ket{g}^{\otimes N} ),
\end{split}
\end{equation}
where the amount of entanglement and relative phase  are obtained from the magnitude and phase, respectively, of the overlap of the clock states evolving in different branches of the superposition $\ket{\tau_\alpha}=\sum_k\exp(-i\tau \hat{h}_{k, \alpha} \otimes \hat{\Pi}_k)^N\ket{+}^{\otimes N}\ket{g}^{\otimes N}$. For a single clock we have
\begin{equation}
\begin{split}
    \langle\tau_L|\tau_R\rangle
={}&
\frac12\left(e^{-i\phi_0}+e^{-i\phi_1}\right) \\
={}&
e^{-i(\phi_0+\phi_1)/2}
\cos\left(\frac{\Delta\phi}{2}\right),
    \end{split}
\end{equation}
with $
\phi_j\simeq
\frac{\tau GM_PM_j}{\hbar}\frac{\Delta x}{d^2}+
\frac{\tau M_j}{2\hbar\omega_j^2}
\left(g_L^2-g_R^2\right)$ and $\Delta\phi=\phi_1-\phi_0\simeq \tau\omega_{c}
\left(
\frac{GM_P}{c^2}\frac{\Delta x}{d^2}
+\frac{g_L^2-g_R^2}{\omega_0^2c^2}
\right)$ where we used that for the two level clock $k=0,1$ and $M_{1}-M_{0}=\hbar\omega_c/c^2$. For $N$ clocks, the amount of entanglement between the particle and clocks are given by $
|\langle\tau_L|\tau_R\rangle|^N$ and for  any individual clock its entanglement is given by $
|{\langle\tau_L|\tau_R\rangle}|^{N-1}$, due to tracing all but one clock. For large number of clocks considered here, we approximate $N-1\approx N$ and thus both the effect on the particle and on any individual clock are exponentially enhanced
\begin{equation}\label{vislos123}
|\left\langle\tau_{L} \mid \tau_{R}\right\rangle|^N\approx\exp\left[
-\frac12
\left(
\frac{\sqrt{N}\,\tau GM_P\omega_{c}\Delta x}{2c^2d^2}
\right)^2\right].
\end{equation}
This expression reproduces the  decoherence time in Eq.~\eqref{tau1234}, as we dropped the term coming from the mass-energy correction to the trap minimum, $\frac{g_L^2-g_R^2}{\omega_0^2c^2}$, which for the parameters considered is negligible ($35$ orders of magnitude smaller than the term we retain). This analysis further justifies the use of the interaction Hamiltonian given in Eq.~\eqref{clockH} in the regime considered in this work.

\textit{Proposed experimental protocol.}  We propose to initialise a solid-state array of nuclear clocks each in the $\ket{+}$ state, and each clock interacts with a delocalised levitated Planck-mass particle. The entanglement of the array of clocks with the levitated particle is accessible through standard population difference measurements on the individual clocks. We first either split the wave-packet of the levitated particle or delocalise its wavepacket, and allow the interaction Hamiltonian of Eq.~\eqref{clockH} to generate entanglement. The most experimentally relevant protocol will involve a gaussian de-localised wavepacket (see Appendix~\ref{smalldelocalisation}), corresponding to the  results discussed in Eq.~\eqref{tau1234} and Eq.~\eqref{tdm}. The levitated particle is then recombined and measured, factoring out the $N$ clocks into a many-body entangled state. Such expansion and recombination with near-unit fidelity was suggested in Ref.~\cite{PhysRevLett.127.023601}, with an expansion coefficient corresponding to a factor of up to $10^5$ times the zero-point-motion, yielding delocalisation length scales on the order of $10 - 100 \;  \mathrm{nm}$, the characteristic de-localisation length scale that we have used through-out this paper.

\textit{Master equation.} In the experimentally relevant regime, we can develop a master equation to describe the decoherence of the levitated particle similar to that done in Ref.~\cite{PikovskiIgor2015Uddt}. The resulting decoherence time of a spatial superposition of the particle is (for distance of the particle to the face of the crystal $D$, with crystal volume $V$)
\begin{equation}\label{tau123}
    \tau_{\mathrm{dec}}=\frac{\sqrt{2} \hbar c^2 }{\sqrt{N} \Delta E_{\mathrm{single}} G M_p \Delta x} \sqrt{\frac{2 D V}{\pi}},  \end{equation}
in full agreement with Eq.~\eqref{tau1234}.  For a single scandium clock with an energy uncertainty of $\Delta E \sim \hbar\omega$ in the state $\ket{+}$, the corresponding decoherence rate is $\Gamma_{\mathrm{dec}} \sim 10^{-17};\mathrm{Hz}$, while for $N = 10^{23}$ clocks this becomes $10^{-6}\;\mathrm{Hz}$. Ref.~\cite{PhysRevLett.127.023601} provides a noise model for a levitated particle, where the calculated bounds apply to a specific expansion--recompression protocol, and may not be directly applicable to the present work. However, there are two relevant requirements: that any  noise still allows for non-vanishing coherence, and that it does not scale with the distance to the clocks and their number, thus allowing the experiment to distinguish the different contributions.

We next construct the decoherence rate of a single clock after tracing out the levitated particle as an environment. We obtain a similar decoherence rate but now given as a pure dephasing term (in the Born-approximation):
\begin{equation}
\begin{split}
    \dot{\rho}_{q}(t) &=  -\frac{N\left(G M_p \right)^2 }{d^2 c^4} \\ &\int_0^t d s\left[\hat{H}_0,\left.\left[\hat{H}_0, \rho_{q}(t-s)\right]\right|_s\right],    
\end{split}
\end{equation}
where $\hat{H}_0$ is the clock Hamiltonian and $\rho_q$ is the density matrix of a qubit that forms the remaining clock sub-system after tracing out the motional degree of freedom of the levitated particle and the $N-1$ clocks. Solving for the decay of the off-diagonal terms of the density matrix (and extending to the standard $\sqrt{N}$ ) we obtain
\begin{equation}
  \rho_{01}(t)=\rho_{01}(0) \cos \left(\frac{2G M_p \sqrt{N} \omega_c}{c^2 d}t .\right)
\end{equation}
for the dephasing of the off-diagonal elements of the $\ket{+}$ state of a single clock in an ensemble of N uncorrelated clocks. For the expected generalisation to the case of a crystal slab in the regime of small de-localisation $\Delta x \ll D$ we replace $d\rightarrow \frac{1}{\Delta x} \sqrt{\frac{2DV}{\pi}}$, hence deriving the previously discussed decoherence rate from a full master equation picture.

\textit{Interpretation.} 
The proposed experiment aims to probe gravitationally mediated entanglement by leveraging interactions between a levitated massive particle and an array of quantum clocks. This experimental proposal transcends prior gravitationally induced entanglement proposals by incorporating relativistic corrections to the gravitational interaction of quantum systems. In doing so, it brings into experimental reach quantised active gravitational mass, here, the internal mass-energy of the clocks. Testing the quantised nature of mass-energies has been explored in the past \cite{ZychMagdalena2018Qfot,PhysRevA.106.052801} and specfically for active gravitational mass, in \cite{fragkos2025probingclassicalquantumviolations, RuizEstebanCastro2017Eoqc}, but remained conceptual in nature. This proposal takes the time scale for this effect from $10^{31} \; \mathrm{s}$ to $10^6 \; \mathrm{s}$, opening the path to conceivable experiments that test for entanglement mediated by the gravitational interaction with quantum clocks. Our work further highlights that introducing  quantized gravitational degrees of freedom is neither required nor implied by gravity induced entanglement, also beyond fully non-relativistic regime considered in the original proposals \cite{bose2017spin, MVedral} and discussed in \cite{Foo2026}.

\textit{Conclusion.}
We proposed a test of gravitationally induced entanglement arising from the interaction between a levitated Planck-mass particle and an array of quantum clocks. While previous works suggested that laboratory tests of gravitational entanglement of quantum clocks would be infeasible because of the clocks’ extremely small mass-energy, we show that this limitation can be overcome by combining the large mass of a levitated particle with a macroscopic array of N quantum clocks. This unlocks a novel path for experiments in which the ability of gravity to entangle can be tested in a relativistic (albeit still low-energy) regime. Such a setting avoids loopholes in the interpretation of the result through non-relativistic Newtonian potentials, which provides a fully classical though non-local explanation for gravity-mediated entanglement. In contrast, our proposed experiment would falsify the non-relativistic Newtonian limit of gravity, where internal energy does not contribute to the system's mass, which predicts no entanglement in our proposed experiment. Finally, we note that such an experiment would access physical regime which has not been tested in any prior experiment -- the interface of quantum theory and gravity beyond the non-relativistic Newtonian limit. This again differs from  gravity mediated entanglement proposals with spatial superpositions of masses interacting via Newtonian potential. 

\section*{Acknowledgements}
We thank Jun Ye for helpful discussions. This material is based upon work supported by the Knut and Alice Wallenberg foundation through a Wallenberg Academy Fellowship No. 2021.0119 and the General Sir John Monash Foundation.
\appendix
\section{Dynamics of a cat state of a particle interacting with an array of clocks}\label{decapp1}
If the clock is initially in the $\ket{+}$ state, time evolution under the interaction Hamiltonian produces entanglement:
\begin{equation}\label{cat1}
  \ket{\psi(0)} = \frac{1}{\sqrt{2}}\left( \ket{-x}\ket{\tau_-}  +  \ket{+x}\ket{\tau_+} \right),
\end{equation}
where the respective clock states are:
\begin{equation}
\begin{split}
    \ket{\tau_- } &= \frac{1}{\sqrt{2}}(\exp \left(+i t \frac{G M}{|d+\Delta x|} \frac{\omega_c}{2 c^2}\right)\ket{g} \\&+  \exp \left(-i t \frac{G M}{\left|d +  \Delta x \right|} \frac{\omega_c }{2c^2}\right)\ket{e}),\\
    \ket{\tau_+ } &= \frac{1}{\sqrt{2}}(\exp \left(+i t \frac{G M}{|d|} \frac{\omega_c}{2 c^2}\right)\ket{g} \\&+  \exp \left(-i t \frac{G M}{\left|d  \right|} \frac{\omega_c }{2c^2}\right)\ket{e}).
\end{split}
\end{equation}
In the regime for which $\Delta x \gg d$, we obtain
\begin{equation}
\begin{split}
    \langle\tau_- | \tau_+ \rangle &= \frac{1}{2}\left(\exp \left(+i t \frac{G M}{|d|} \frac{\omega_c}{2 c^2}\right) + \exp(-it\frac{GM\omega_c}{2dc^2}) \right) \\
    &= \cos\left(t \frac{G M}{|d|} \frac{\omega_c}{2 c^2}\right),
\end{split}
\end{equation}
which for short times such that $t \ll \left( \frac{G M}{|d|} \frac{\omega_c}{2 c^2} \right)^{-1}$, this can be expanded as 
\begin{equation}
\begin{split}
    \cos\left(t \frac{G M}{|d|} \frac{\omega_c}{2 c^2}\right) &\approx 1 - \left(t \frac{G M}{|d|} \frac{\omega_c}{2 c^2}\right)^2 \\ &\approx \exp\left(- \left(t \frac{G M}{|d|} \frac{\omega_c}{2 c^2}\right)^2\right),
\end{split}
\end{equation}
and therefore associated with a decoherence time-scale for the decay of the off-diagonal terms as 
\begin{equation}
    t_d=\frac{d}{G M} \frac{2 c^2}{\omega_c}.
\end{equation}
In the regime $\Delta x\ll d$, the overlap of the clock states is
\begin{equation}
    \langle\tau_-|\tau_+\rangle
    \simeq \cos\left(
    t\frac{GM\omega_c\Delta x}{2c^2d^2}\right)\simeq\exp\left(-\left(t\frac{GM\omega_c\Delta x}{2c^2d^2}\right)^2\right).
\end{equation}
The corresponding decoherence time is
\begin{equation}\label{eq:appendix-smalldeloc}
t_d=\frac{2c^2d^2}{GM\omega_c\Delta x}.
\end{equation}

To see the enhancement with the array of N initially uncorrelated clocks initialised in the $\ket{+}$ state, we note that this modifies the quantum state in Eq.~\eqref{cat1} to be 
\begin{equation}
    \ket{\psi(0)} = \frac{1}{\sqrt{2}}\left( \ket{-x}\ket{\tau_-}^{\otimes N}  +  \ket{+x}\ket{\tau_+}^{\otimes N} \right),
\end{equation}
and therefore the visibility loss becomes for each individual clock becomes
\begin{equation}
\begin{split}
    \left\langle\tau_{-} \mid \tau_{+}\right\rangle^{N-1} &= \cos \left(t \frac{G M}{|d|} \frac{\omega_c}{2 c^2}\right)^{N-1} \\
    &\approx 1-(N -1)\left(t \frac{G M}{|d|} \frac{\omega_c}{2 c^2}\right)^2 \\
    &\approx 1-\left(t \sqrt{N} \frac{G M}{|d|} \frac{\omega_c}{2 c^2}\right)^2,
\end{split}
\end{equation}
where in the last line we have used that $N - 1 \approx N$ for large N. Therefore, for interactions with an array of N uncorrelated clocks, the decoherence time-scale becomes
\begin{equation}
    t_d=\frac{d}{G M} \frac{2 c^2}{\sqrt{N}\omega_c}.
\end{equation}
Similary for the $\Delta x \ll d$ the multi-clock decoherence time scale is
\begin{equation}
    t_d=\frac{2d^2c^2}{G M\sqrt{N}\omega_c\Delta x}.
\end{equation}

\section{Dynamics of a de-localised massive particle interacting with an array of clocks}\label{smalldelocalisation}
We now consider corrections for the experimentally accessible regime in which the delocalisation of the levitated particle is much smaller than the interaction distance to the array of clocks. In this way, we consider a levitated particle in some gaussian-wavepacket:
\begin{equation}
   \ket{\psi_p(0)} = \sqrt[4]{\frac{1}{\pi \sigma^2}} \int \mathrm{~d} x^{\prime} \exp \left(-\frac{\left(x^{\prime}-r\right)^2}{2 \sigma^2}\right)\left|x^{\prime}\right\rangle,
\end{equation}
and consider that the spatial extent of the wave-packet's position $\sigma$ is much smaller than the distance to each of the clocks: $\sigma \ll d$. In this case, the clock state will become entangled with the position of the delocalised source mass
\begin{equation}
    \ket{\psi(t)} = \sqrt[4]{\frac{1}{\pi \sigma^2}} \int \mathrm{~d} x^{\prime} \exp \left(-\frac{\left(x^{\prime}-r\right)^2}{2 \sigma^2}\right)\left|x^{\prime}\right\rangle\left|\tau_{x^{\prime}}\right\rangle,
\end{equation}
where 
\begin{equation}
\begin{aligned}
\left|\tau_{x^{\prime}}\right\rangle & =\frac{1}{\sqrt{2}}\left(\exp \left(+i t \frac{G M}{|d + x^{\prime}|} \frac{\omega_c}{2 c^2}\right)|g\rangle\right. \\
& \left.+\exp \left(-i t \frac{G M}{|d + x^{\prime}|} \frac{\omega_c}{2 c^2}\right)|e\rangle\right).
\end{aligned}
\end{equation}
For an array of N-clocks this becomes the many-clock entangled state
\begin{equation}
    \ket{\psi(t)} = \sqrt[4]{\frac{1}{\pi \sigma^2}} \int \mathrm{~d} x^{\prime} \exp \left(-\frac{\left(x^{\prime}-r\right)^2}{2 \sigma^2}\right)\left|x^{\prime}\right\rangle\left|\tau_{x^{\prime}}\right\rangle^{\otimes N}.
\end{equation}
The overlap,
\begin{equation}
  \langle \tau_{x^{\prime \prime}}  \left|\tau_{x^{\prime}}\right\rangle = \cos\left( \frac{itGM\omega_c}{2c^2} \left( \frac{1}{d + x^\prime} - \frac{1}{d + x^{\prime\prime}}  \right)  \right),
\end{equation} 
will quantify the extent of the visibility loss. The extent of the visibility loss can therefore be computed as
\begin{equation}
\begin{split}
	\mathcal{V}  &=   \sqrt[2]{\frac{1}{\pi \sigma^2}} \int \mathrm{~d} x^{\prime}\int \mathrm{~d} x^{\prime\prime} \exp \left(-\frac{\left(x^{\prime}-r\right)^2}{2 \sigma^2}\right) \\
	  &\exp \left(-\frac{\left(x^{\prime\prime}-r\right)^2}{2 \sigma^2}\right)\langle \tau_{x^{\prime \prime}}  \left|\tau_{x^{\prime}}\right\rangle^{N-1},
\end{split}
\end{equation} 
reproducing the same earlier derived rates in the relevant regimes.  

In the regime where the spatial extent of the wavepacket
is much smaller than the distance to the clocks, $\Delta x\ll d$, we can expand
\begin{equation}
\frac{1}{d+x} \simeq \frac{1}{d}-\frac{x}{d^2}.
\end{equation}
Consequently,
\begin{equation}
\frac{1}{d+x'}-\frac{1}{d+x''}\simeq -\frac{x'-x''}{d^2}.
\end{equation}
The overlap therefore becomes
\begin{equation}
\left\langle\tau_{x''}|\tau_{x'}\right\rangle \simeq \cos\left[\frac{GM_p\omega_i t}{2c^2d^2}(x'-x'')\right].
\end{equation}
For sufficiently short times, we use the approximation
\begin{equation}
\cos\theta\simeq 1-\frac{\theta^2}{2},
\end{equation}
which gives
\begin{equation}
\cos^{N-1}\theta \simeq \exp\left[-\frac{N-1}{2}\theta^2\right].
\end{equation}
Substituting into the visibility, we obtain
\begin{equation}
\begin{aligned}
\mathcal{V}(t) &\propto \int dx'\int dx'' \exp\left[-\frac{(x'-r)^2+(x''-r)^2}{2\sigma^2}\right]
\\
&\quad\times
\exp\left[-\frac{(N-1)G^2M_p^2\omega_i^2t^2}{8c^4d^4}(x'-x'')^2\right].
\end{aligned}
\end{equation}
To evaluate the Gaussian integrals, introduce the variables
\begin{equation}
    u=\frac{x'+x''}{2}-r,
    \qquad v=x'-x''.
\end{equation}
Under this transformation,
\begin{equation}
    (x'-r)^2+(x''-r)^2 = 2u^2+\frac{v^2}{2},
\end{equation}
and the Jacobian is unity. The visibility can therefore be written as
\begin{equation}
\begin{aligned}
\mathcal{V}(t) &\propto \int du\, \exp\left(-\frac{u^2}{\sigma^2}\right)
\\ &\quad\times \int dv\, \exp\left[-\left(\frac{1}{4\sigma^2}+\frac{(N-1)G^2M_p^2\omega_i^2t^2}{8c^4d^4}\right)v^2\right].
\end{aligned}
\end{equation}
Using the Gaussian integral
\begin{equation}
\int_{-\infty}^{\infty}dx\,e^{-ax^2}=\sqrt{\frac{\pi}{a}},
\end{equation}
and normalising the visibility such that $\mathcal{V}(0)=1$, we find
\begin{equation}
\mathcal{V}(t)\simeq\left[1+\frac{(N-1)G^2M_p^2\omega_i^2\sigma^2t^2}{2c^4d^4}\right]^{-1/2}.
\end{equation}
For short times, using
\begin{equation}
(1+\epsilon)^{-1/2}\simeq \exp\left(-\frac{\epsilon}{2}\right),
\end{equation}
we obtain
\begin{equation}
\mathcal{V}(t)\simeq \exp\left[-\frac{(N-1)G^2M_p^2\omega_i^2\sigma^2t^2}{4c^4d^4}\right].
\end{equation}
We define the characteristic decoherence time through
\begin{equation}
    \mathcal{V}(t)\simeq\exp\left[-\left(\frac{t}{t_{\mathrm{dec}}}\right)^2\right].
\end{equation}
Comparing the exponents gives
\begin{equation}
\frac{1}{t_{\mathrm{dec}}^2}=\frac{(N-1)G^2M_p^2\omega_i^2\sigma^2}{4c^4d^4}.
\end{equation}
Thus,
\begin{equation}
t_{\mathrm{dec}}=\frac{2c^2d^2}{GM_p\sqrt{N-1}\,\omega_i\sigma}.
\end{equation}
Finally, identifying the spatial width of the wavepacket as $\Delta x=\sigma$, and taking the large-$N$ limit, $\sqrt{N-1}\simeq\sqrt{N}$, we recover the result of Eq.~\eqref{eq:appendix-smalldeloc}
\begin{equation}
    t_{\mathrm{mass}-N-\mathrm{clocks}}=\frac{2c^2d^2}{GM_p\sqrt{N}\,\omega_i\Delta x}.
\end{equation}

\section{Dynamics of the master equation}
If we examine the master equation for the reduced state of a single clock with $\hat{H}_0 = \frac{\omega_c\hat{\sigma_z}}{2}$:
\begin{equation}
\begin{split}
        \dot{\rho}_q(t) &= -\frac{\left(\sqrt{N}G M_p\right)^2}{d^2 c^4} \int_0^t d s\left[\hat{H}_0,\left.\left[\hat{H}_0, \rho_q(t-s)\right]\right|_s\right]   \\
        &= -\frac{\left(\sqrt{N} \omega_c G M_p\right)^2}{d^2 c^4}\int_0^t d s \left(\begin{array}{cc}
0 & 4 p_{01}(t-s) \\
4 p_{10}(t-s) & 0
\end{array}\right),
\end{split}
\end{equation}
which after a change of variables to $u = t-s$ and differentiating both sides reduces to a second order differential equation for the off-diagonal elements:
\begin{equation}
    \ddot{p_{01}}(t) = -4\left(\kappa\right)^2 p_{01}(t),
\end{equation}
with $\kappa = \frac{\left(\sqrt{N} \omega_c G M_p\right)}{d c^2}$. Which has the straight-forward analytical solution
\begin{equation}
    \rho_{01}(t)=\rho_{01}(0) \cos \left(2 \kappa t\right).
\end{equation}

\section{Scaling with distances}\label{crystal123}
The prior modeling considered if all the clocks are at approximately the same distance from the levitated particle. In this section we now examine corrections to that approximation. 

\subsection{Crystal}
If the levitated particle is some fixed distance $D$ from the centre of a crystal slab, in the regime for which the particle delocalisation $\Delta x$ is much smaller than the distance to the crystal slab (with crystal charateristic size)
\begin{equation}
    t_{\text {finite}, \text { small }} \approx \frac{2 c^2 D^2}{G M_p \omega_c \sqrt{N} \Delta x}.
\end{equation}
For a Planck-mass source positioned a distance $D$ from

the surface of a scandium crystal, the decoherence time

in the regime $\Delta x \ll D \ll L$ is

\begin{equation}
t_{\mathrm{finite}} \approx \frac{2c^2}{G M_p \omega_c \Delta x} \sqrt{\frac{2D}{\pi n}},
\end{equation}
where $n$ is the nuclear clock number density and $L$ is the characteristic crystal size. We consider the parameters $M_p = 2.18\times10^{-8}\,\mathrm{kg}$, $D = 10^{-6}\,\mathrm{m}$, $\omega_c = 2\pi\times10^{14}\,\mathrm{s^{-1}}$, $\Delta x = 10^{-7}\,\mathrm{m}$, $n = 4\times10^{28}\,\mathrm{m^{-3}},
\mathrm{m^3\,kg^{-1}\,s^{-2}}$. For a centimetre-scale crystal, the conditions $\Delta x \ll D \ll L$ are satisfied. Substituting these values gives,
\begin{equation}
t_{\mathrm{finite}}\approx 6\times 10^6\,\mathrm{s}.
\end{equation}

\subsection{Concentric rings}
The number of atoms, in an optical lattice, on the surface of a sphere of radius $r$ is $N = \frac{4 \pi r^2 }{\lambda^2}$. Therefore, for a series of concentric spheres surrounding the levitated particle, the time evolved state is:
\begin{equation}
    \begin{split}
 \ket{\psi(t)} &= \sqrt[4]{\frac{1}{\pi \sigma^2}} \int \mathrm{~d} x^{\prime} \exp \left(-\frac{\left(x^{\prime}-r\right)^2}{2 \sigma^2}\right) \\& \frac{1}{\sqrt{2}}(\prod_a\exp \left(+i t \frac{4\pi}{\lambda^2} \frac{G M (10^{-4} + a\lambda)^2}{\left|10^{-4} + a\lambda\right|} \frac{\omega_c}{c^2}\right)\ket{g}^{\otimes N} \\&+ \prod_a\exp \left(-i t \frac{4\pi}{\lambda^2} \frac{G M(10^{-4} + a\lambda)^2}{\left|10^{-4} +a\lambda\right|} \frac{\omega_c}{c^2}\right)\ket{e}^{\otimes N})\left|x^{\prime}\right\rangle,
 \end{split}
\end{equation}
the $10^{-4}$ term, will give the same scaling with $K$ as in the previous section, however, the second term, which results in a sum over the number of concentric spheres, will yield $\sum_a a = \frac{1}{2}K\left(K + 1\right)$, giving back an $N$-scaling in the phase:
\begin{equation}
    \begin{split}
 \ket{\psi(t)} &= \sqrt[4]{\frac{1}{\pi \sigma^2}} \int \mathrm{~d} x^{\prime} \exp (-\frac{\left(x^{\prime}-r\right)^2}{2 \sigma^2}) \\ \frac{1}{\sqrt{2}}&(\exp (+i t \frac{4\pi G M 10^{-4} K}{\lambda^2} \\&+  \frac{4\pi G M (K(K+1))}{\lambda} \frac{\omega_c}{c^2})\ket{g}^{\otimes N} \\+ \exp &(-i t \frac{4\pi G M 10^{-4} K}{\lambda^2} \\&+  \frac{4\pi G M (K(K+1))}{\lambda} \frac{\omega_c}{c^2})\ket{e}^{\otimes N})\left|x^{\prime}\right\rangle.
 \end{split}
\end{equation}
Now the number of atoms in $K$ concentric circles scales as $(K + 1)K^2$. Therefore, the $K(K +1)$ term above gives approximately an $N^{2/3}$ scaling.

\section{NO-ON state enhancement.} 
In atomic clock set-ups, the decoherence time-scale can be reduced by a further factor of $\sqrt{N}$ if an initial entangled state of the internal energies of the clocks is used. In order to see this, we consider an initial state of the systems 
\begin{equation}
    \left|\psi_c(0)\right\rangle=\frac{1}{2}\left(|g\rangle^{\otimes N}+|e\rangle^{\otimes N}\right)(|-x\rangle+|+x\rangle),
\end{equation}
the dynamics under the gravitational interaction produces a $N$-factor enhancement of the relative phase 
 \begin{equation}
  \begin{split}
    \ket{\psi(t)} &= \frac{1}{2}(\ket{-x}\left(\ket{g}^{\otimes N} + \ket{e}^{\otimes N} \right)\\
    &+\frac{1}{2}(\ket{+x}(\exp \left(+i t \frac{G M N}{|d|} \frac{\omega_c}{2 c^2}\right)\ket{g}^{\otimes N} \\&+ \exp \left(-i t \frac{G M N}{|d|} \frac{\omega_c}{2 c^2}\right)\ket{e}^{\otimes N} ).
  \end{split}
\end{equation} 
After this gravitational interaction for time $t$, the application of $N-1$, C-NOT gates, can now move the phase information on to a single clock:
 \begin{equation}
  \begin{split}
    \ket{\psi(t)} &= \frac{1}{2}(\ket{-x}\left(\ket{g} + \ket{e}\right)\ket{g}^{N-1}\\
    &+\frac{1}{2}(\ket{+x}(\exp \left(+i t \frac{G M N}{|d|} \frac{\omega_c}{2 c^2}\right)\ket{g} \\&+ \exp \left(-i t \frac{G M N}{|d|} \frac{\omega_c}{2 c^2}\right)\ket{e} )\ket{g}^{N-1},
  \end{split}
\end{equation} 
tracing out the particle and making measurements on the remaining clock in the $\ket{+},\ket{-}$ basis that produces visibility drops according to $\left\langle\tau_{-} \mid \tau_{+}\right\rangle = \cos \left(t \frac{G M N \omega_c}{2d c^2}\right)$, corresponding to a decoherence time-scale of 
\begin{equation}
    t_d = \frac{2d c^2}{G M N \omega_c},
\end{equation} 
improving upon the time-scale of Eq.~\eqref{td1} by a factor of $\sqrt{N}$. For an atomic clock, with $\frac{\omega}{2\pi} = 10^{14} \; \mathrm{Hz}$, and $N = 10^{6}$, we obtain: $t_d \sim 10^{8} \; \mathrm{s}$, and $N = 10^{23}$, we obtain: $t_d \sim 10^{-9} \; \mathrm{s}$. Now, with the generalisation to the crystal slab decoherence time-scale, undertaken by replacing $d\rightarrow \frac{1}{\Delta x} \sqrt{\frac{2DV}{\pi}}$, this gives $t_d = 10^{15} \; \mathrm{s}$ and $t_d  = 10^{-3} \; \mathrm{s}$ respectively. However, we don't consider such states central to witnessing the effect, because while amplifying the entanglement, they also amplify the clock state linewidth in the presence of gravitational interactions as highlighted in Ref.~\cite{beyer2026gravitationalinteractionsnuclearclocks}. For this reason we retain the initial state being $N$ uncorrelated $\ket{+}$ states for the clocks.

\bibliography{refs}

@CONTROL{REVTEX41Control}

@CONTROL{apsrev41Control,author="08",editor="1",pages="1",title="0",year="1"}

@article{Fortier:26,
author = {Tara M. Fortier and Andre N. Luiten and Helen S. Margolis},
journal = {Optica},
number = {1},
pages = {143--163},
publisher = {Optica Publishing Group},
title = {Optical atomic clocks: defining the future of time\&\#x00A0;and frequency metrology},
volume = {13},
month = {Jan},
year = {2026},
url = {https://opg.optica.org/optica/abstract.cfm?URI=optica-13-1-143},
doi = {10.1364/OPTICA.575770}
}

@article{arnold2026lu,
  author    = {Arnold, K. J. and Lee, M. D. K. and Zhao, Qi and Qin, Qichen and Zhang, Zhao and Jayjong, N. and Barrett, M. D.},
  title     = {Lu$^{+}$ optical frequency references with accuracy verified at the 19th digit},
  journal   = {Nature},
  volume    = {658},
  number    = {8134},
  pages     = {60--64},
  year      = {2026},
  month     = {Oct},
  doi       = {10.1038/s41586-026-11072-8},
  publisher = {Nature Publishing Group}
}

@article{PhysRev.175.782,
  title = {Correlation between Mossbauer Resonance Strength and Second-Order Doppler Shift: Estimate of Zero-Point Velocity},
  author = {Taylor, R. Dean and Craig, Paul P.},
  journal = {Phys. Rev.},
  volume = {175},
  issue = {3},
  pages = {782--787},
  numpages = {0},
  year = {1968},
  month = {Nov},
  publisher = {American Physical Society},
  doi = {10.1103/PhysRev.175.782},
  url = {https://link.aps.org/doi/10.1103/PhysRev.175.782}
}

@article{Foo2026,
  author = {Foo, Joshua and Suryaatmadja, Cendikiawan and Mann, Robert B. and Zych, Magdalena},
  title = {Relativity and decoherence of spacetime superpositions},
  journal = {npj Quantum Information},
  year = {2026},
  volume = {12},
  number = {1},
  pages = {116},
  doi = {10.1038/s41534-026-01234-x},
  url = {https://doi.org/10.1038/s41534-026-01234-x}
}

@article{RuizEstebanCastro2017Eoqc,
issn = {0027-8424},
journal = {Proceedings of the National Academy of Sciences - PNAS},
pages = {E2303--E2309},
volume = {114},
publisher = {National Academy of Sciences},
number = {12},
year = {2017},
title = {Entanglement of quantum clocks through gravity},
copyright = {Volumes 1–89 and 106–114, copyright as a collective work only; author(s) retains copyright to individual articles},
address = {United States},
author = {Ruiz, Esteban Castro and Giacomini, Flaminia and Brukner, Caslav},
}

@article{PikovskiIgor2015Uddt,
issn = {1745-2473},
journal = {Nature physics},
number = {8},
pages = {668-672},
title = {Universal decoherence due to gravitational time dilation},
volume = {11},
year = {2015},
author = {Pikovski, Igor and Zych, Magdalena and Costa, Fabio and Brukner, Caslav},
}

@misc{fragkos2025probingclassicalquantumviolations,
      title={Probing classical and quantum violations of the equivalence of active and passive gravitational mass}, 
      author={Vasileios Fragkos and Igor Pikovski},
      year={2025},
      eprint={2503.16198},
      archivePrefix={arXiv},
      primaryClass={quant-ph},
      url={https://arxiv.org/abs/2503.16198}, 
}

@misc{alma991014454786403131,
author = {Kiefer, Claus.},
address = {Oxford},
booktitle = {Quantum gravity},
edition = {3rd ed},
isbn = {0-19-162885-9},
publisher = {Oxford University Press},
series = {International series of monographs on physics  Quantum gravity},
title = {Quantum gravity },
year = {2012},
}

@article{PhysRevA.88.022318,
  title = {Quantum computations without definite causal structure},
  author = {Chiribella, Giulio and D'Ariano, Giacomo Mauro and Perinotti, Paolo and Valiron, Benoit},
  journal = {Phys. Rev. A},
  volume = {88},
  issue = {2},
  pages = {022318},
  numpages = {15},
  year = {2013},
  month = {Aug},
  publisher = {American Physical Society},
  doi = {10.1103/PhysRevA.88.022318},
  url = {https://link.aps.org/doi/10.1103/PhysRevA.88.022318}
}

@article{OreshkovOgnyan2012Qcwn,
author = {Oreshkov, Ognyan and Costa, Fabio and Brukner, Caslav},
address = {London},
copyright = {The Author(s) 2012},
issn = {2041-1723},
journal = {Nature communications},
number = {1},
pages = {1092-},
publisher = {Nature Publishing Group UK},
title = {Quantum correlations with no causal order},
volume = {3},
year = {2012},
}

@article{CarlessoMatteo2022Psaf,
author = {Carlesso, Matteo and Donadi, Sandro and Ferialdi, Luca and Paternostro, Mauro and Ulbricht, Hendrik and Bassi, Angelo},
address = {London},
copyright = {Springer Nature Limited 2022},
issn = {1745-2473},
journal = {Nature physics},
number = {3},
pages = {243-250},
publisher = {Nature Publishing Group UK},
title = {Present status and future challenges of non-interferometric tests of collapse models},
volume = {18},
year = {2022},
}

@misc{beyer2026gravitationalinteractionsnuclearclocks,
      title={Gravitational Interactions in Nuclear Clocks}, 
      author={Konstantin Beyer and Niklas E. Önne and Shimon Kolkowitz and Dolev Bluvstein and Mikhail Lukin and Jun Ye and Igor Pikovski},
      year={2026},
      eprint={2610.38832},
      archivePrefix={arXiv},
      primaryClass={quant-ph},
      url={https://arxiv.org/abs/2610.38832}, 
}

@article{hl1c-t8z9,
  title = {Quantum Effects in Gravity Beyond the Newton Potential from a Delocalized Quantum Source},
  author = {Chen, Lin-Qing and Giacomini, Flaminia},
  journal = {Phys. Rev. X},
  volume = {15},
  issue = {3},
  pages = {031063},
  numpages = {19},
  year = {2025},
  month = {Sep},
  publisher = {American Physical Society},
  doi = {10.1103/hl1c-t8z9},
  url = {https://link.aps.org/doi/10.1103/hl1c-t8z9}
}

@article{PhysRevLett.127.023601,
  title = {Large Quantum Delocalization of a Levitated Nanoparticle Using Optimal Control: Applications for Force Sensing and Entangling via Weak Forces},
  author = {Weiss, T. and Roda-Llordes, M. and Torrontegui, E. and Aspelmeyer, M. and Romero-Isart, O.},
  journal = {Phys. Rev. Lett.},
  volume = {127},
  issue = {2},
  pages = {023601},
  numpages = {6},
  year = {2021},
  month = {Jul},
  publisher = {American Physical Society},
  doi = {10.1103/PhysRevLett.127.023601},
  url = {https://link.aps.org/doi/10.1103/PhysRevLett.127.023601}
}

@misc{haustein2019massenergyequivalenceharmonicallytrapped,
      title={Mass-energy equivalence in harmonically trapped particles}, 
      author={Rebecca Haustein and Gerard J. Milburn and Magdalena Zych},
      year={2019},
      eprint={1906.03980},
      archivePrefix={arXiv},
      primaryClass={quant-ph},
      url={https://arxiv.org/abs/1906.03980}, 
}

@article{RevModPhys.97.015003,
  title = {Massive quantum systems as interfaces of quantum mechanics and gravity},
  author = {Bose, Sougato and Fuentes, Ivette and Geraci, Andrew A. and Khan, Saba Mehsar and Qvarfort, Sofia and Rademacher, Markus and Rashid, Muddassar and Toro\ifmmode \check{s}\else \v{s}\fi{}, Marko and Ulbricht, Hendrik and Wanjura, Clara C.},
  journal = {Rev. Mod. Phys.},
  volume = {97},
  issue = {1},
  pages = {015003},
  numpages = {71},
  year = {2025},
  month = {Feb},
  publisher = {American Physical Society},
  doi = {10.1103/RevModPhys.97.015003},
  url = {https://link.aps.org/doi/10.1103/RevModPhys.97.015003}
}

@article{PhysRevD.108.L101702,
  title = {What gravity mediated entanglement can really tell us about quantum gravity},
  author = {Mart\'{\i}n-Mart\'{\i}nez, Eduardo and Perche, T. Rick},
  journal = {Phys. Rev. D},
  volume = {108},
  issue = {10},
  pages = {L101702},
  numpages = {7},
  year = {2023},
  month = {Nov},
  publisher = {American Physical Society},
  doi = {10.1103/PhysRevD.108.L101702},
  url = {https://link.aps.org/doi/10.1103/PhysRevD.108.L101702}
}

@article{Shvyd,
issn = {0028-0836},
journal = {Nature (London)},
pages = {471--475},
volume = {622},
publisher = {Nature Publishing Group UK},
number = {7983},
year = {2023},
title = {Resonant X-ray excitation of the nuclear clock isomer 45Sc},
copyright = {UChicago Argonne, LLC, Operator of Argonne National Laboratory 2023},
address = {London},
author = {Shvyd’ko, Yuri and Röhlsberger, Ralf and Kocharovskaya, Olga and Evers, Jörg and Geloni, Gianluca Aldo and Liu, Peifan and Shu, Deming and Miceli, Antonino and Stone, Brandon and Hippler, Willi and Marx-Glowna, Berit and Uschmann, Ingo and Loetzsch, Robert and Leupold, Olaf and Wille, Hans-Christian and Sergeev, Ilya and Gerharz, Miriam and Zhang, Xiwen and Grech, Christian and Guetg, Marc and Kocharyan, Vitali and Kujala, Naresh and Liu, Shan and Qin, Weilun and Zozulya, Alexey and Hallmann, Jörg and Boesenberg, Ulrike and Jo, Wonhyuk and Möller, Johannes and Rodriguez-Fernandez, Angel and Youssef, Mohamed and Madsen, Anders and Kolodziej, Tomasz},
organization = {Argonne National Laboratory (ANL), Argonne, IL (United States)},
}

@article{tobar2023detecting,
issn = {2041-1723},
journal = {Nature communications},
pages = {7229--14},
volume = {15},
publisher = {Nature Publishing Group},
number = {1},
year = {2024},
title = {Detecting single gravitons with quantum sensing},
copyright = {The Author(s) 2024. This work is published under http://creativecommons.org/licenses/by/4.0/ (the “License”). Notwithstanding the ProQuest Terms and Conditions, you may use this content in accordance with the terms of the License.},
address = {London},
author = {Tobar, Germain and Manikandan, Sreenath K and Beitel, Thomas and Pikovski, Igor},
}

@INCOLLECTION{zehfeynman,
       author = {{Aspelmeyer}, Markus},
        title = "{When Zeh Meets Feynman: How to Avoid the Appearance of a Classical World in Gravity Experiments}",
    booktitle = {From Quantum to Classical; Essays in Honour of H.-Dieter Zeh},
         year = 2022,
       editor = {{Kiefer}, Claus},
        pages = {85-95}
}

@article{manikandan2024detectingacoherenceradiationfields,
  title = {Testing the coherent-state description of radiation fields},
  author = {Manikandan, Sreenath K. and Wilczek, Frank},
  journal = {Phys. Rev. A},
  volume = {111},
  issue = {3},
  pages = {033705},
  numpages = {12},
  year = {2025},
  month = {Mar},
  publisher = {American Physical Society},
  doi = {10.1103/PhysRevA.111.033705},
  url = {https://link.aps.org/doi/10.1103/PhysRevA.111.033705}
}

@article{MVedral,
  title = {Gravitationally Induced Entanglement between Two Massive Particles is Sufficient Evidence of Quantum Effects in Gravity},
  author = {Marletto, C. and Vedral, V.},
  journal = {Phys. Rev. Lett.},
  volume = {119},
  issue = {24},
  pages = {240402},
  numpages = {5},
  year = {2017},
  month = {Dec},
  publisher = {American Physical Society},
  doi = {10.1103/PhysRevLett.119.240402}
}

@article{PhysRevLett.134.093201,
  title = {Exploring the Dynamical Interplay between Mass-Energy Equivalence, Interactions, and Entanglement in an Optical Lattice Clock},
  author = {Chu, Anjun and Mart\'{\i}nez-Lahuerta, Victor J. and Miklos, Maya and Kim, Kyungtae and Zoller, Peter and Hammerer, Klemens and Ye, Jun and Rey, Ana Maria},
  journal = {Phys. Rev. Lett.},
  volume = {134},
  issue = {9},
  pages = {093201},
  numpages = {9},
  year = {2025},
  month = {Mar},
  publisher = {American Physical Society},
  doi = {10.1103/PhysRevLett.134.093201},
  url = {https://link.aps.org/doi/10.1103/PhysRevLett.134.093201}
}

@article{PhysRevA.106.052801,
  title = {Mass-energy equivalence in gravitationally bound quantum states of the neutron},
  author = {Tobar, Germain and Haine, Simon and Costa, Fabio and Zych, Magdalena},
  journal = {Phys. Rev. A},
  volume = {106},
  issue = {5},
  pages = {052801},
  numpages = {11},
  year = {2022},
  month = {Nov},
  publisher = {American Physical Society},
  doi = {10.1103/PhysRevA.106.052801},
  url = {https://link.aps.org/doi/10.1103/PhysRevA.106.052801}
}

@article{ZychMagdalena2018Qfot,
copyright = {The Author(s) 2018},
issn = {1745-2473},
journal = {Nature physics},
number = {10},
pages = {1027-1031},
publisher = {Nature Publishing Group UK},
title = {Quantum formulation of the Einstein equivalence principle},
volume = {14},
year = {2018},
author = {Zych, Magdalena and Brukner, Caslav},
address = {London},
}

@article{Guerreiro2022quantumsignaturesin,
  title = {Quantum signatures in nonlinear gravitational waves},
  author = {Guerreiro, Thiago and Coradeschi, Francesco and Frassino, Antonia Micol and West, Jennifer Rittenhouse and Schioppa, Enrico Junior},
  journal = {{Quantum}},
  issn = {2521-327X},
  publisher = {{Verein zur F{\"{o}}rderung des Open Access Publizierens in den Quantenwissenschaften}},
  volume = {6},
  pages = {879},
  month = dec,
  year = {2022}
}

@article{PhysRevD.98.124006,
  title = {Quantum interactions between a laser interferometer and gravitational waves},
  author = {Pang, Belinda and Chen, Yanbei},
  journal = {Phys. Rev. D},
  volume = {98},
  issue = {12},
  pages = {124006},
  numpages = {13},
  year = {2018},
  month = {Dec},
  publisher = {American Physical Society}
}

@article{MarshmanGIE,
  title = {Locality and entanglement in table-top testing of the quantum nature of linearized gravity},
  author = {Marshman, Ryan J. and Mazumdar, Anupam and Bose, Sougato},
  journal = {Phys. Rev. A},
  volume = {101},
  issue = {5},
  pages = {052110},
  numpages = {13},
  year = {2020},
  month = {May},
  publisher = {American Physical Society},
  doi = {10.1103/PhysRevA.101.052110},
}

@article{blencowe2013effective,
  title={Effective field theory approach to gravitationally induced decoherence},
  author={Blencowe, MP},
  journal={Physical Review Letters},
  volume={111},
  number={2},
  pages={021302},
  year={2013},
  publisher={APS}
}

@article{kanno2021noise,
  title={Noise and decoherence induced by gravitons},
  author={Kanno, Sugumi and Soda, Jiro and Tokuda, Junsei},
  journal={Physical Review D},
  volume={103},
  number={4},
  pages={044017},
  year={2021},
  publisher={APS}
}

@article{parikh2020noise,
  title={The noise of gravitons},
  author={Parikh, Maulik and Wilczek, Frank and Zahariade, George},
  journal={International Journal of Modern Physics D},
  volume={29},
  number={14},
  pages={2042001},
  year={2020},
  publisher={World Scientific}
}

@article{bose2017spin,
  title={Spin entanglement witness for quantum gravity},
  author={Bose, Sougato and Mazumdar, Anupam and Morley, Gavin W and Ulbricht, Hendrik and Toro{\v{s}}, Marko and Paternostro, Mauro and Geraci, Andrew A and Barker, Peter F and Kim, MS and Milburn, Gerard},
  journal={Physical Review Letters},
  volume={119},
  number={24},
  pages={240401},
  year={2017},
  publisher={APS}
}

@article{Carney_2019,
doi = {10.1088/1361-6382/aaf9ca},
url = {https://dx.doi.org/10.1088/1361-6382/aaf9ca},
year = {2019},
month = {jan},
publisher = {IOP Publishing},
volume = {36},
number = {3},
pages = {034001},
author = {Carney, Daniel and Stamp, Philip C E and Taylor, Jacob M},
title = {Tabletop experiments for quantum gravity: a user’s manual},
journal = {Classical and Quantum Gravity}
}

@book{CécileM.DeWitted.2011TRoG,
series = {Sources 5: Max Planck Research Library for the History and Development of Knowledge},
publisher = {Edition Open Access},
isbn = {3869319631},
year = {2011},
title = {The Role of Gravitation in Physics: Report from the 1957 Chapel Hill Conference},
author = {Cécile M. DeWitt (ed.) and Dean Rickles (ed.)},
}

\end{document}